\documentclass[prb,twocolumn,showpacs,floatfix]{revtex4}
\usepackage{graphicx}
\usepackage{dcolumn}
\usepackage{bm}
\begin{document}
\title{Negative thermal expansion in ZnF$_2$}
\author{Tapan Chatterji, Mohamed Zbiri,  and  Thomas C. Hansen }
\address{Institut Laue-Langevin, B.P. 156, 38042 Grenoble Cedex 9, France\\}
\date{\today}
\begin{abstract}
We have investigated temperature dependence of the lattice parameters and the unit cell volume of ZnF$_2$ by neutron diffraction and have discovered negative thermal expansion (NTE) at low temperature. To understand why this simple compound exhibits NTE we performed first principle calculations. These calculations reproduce qualitatively the experimental temperature dependence of volume. 
\end{abstract}
\maketitle
The negative thermal expansion (NTE) in solids has attracted the renewed attention of condensed matter scientists ever since Sleight and coworkers \cite{mary96,evans96} discovered that ZrW$_2$O$_8$ contracts over a wide temperature range of more than 1000 K. There are excellent review articles \cite{sleight98,evans99}on the NTE of this type of so-called \emph{framework} materials. However NTE is known and has been studied experimentally and theoretically for a long time. Among these Si and Ge and other tetrahedrally bonded crystals at low temperature are classic examples \cite{gibbons58,sparks67,biernacki89}. The NTE is however limited in the low temperature range in these materials, whereas in higher temperature range they exhibit normal positive thermal expansion. There exists a more general review article \cite{barrera05} covering all types of materials that exhibit NTE. Here we report observation of NTE in diamagnetic ZnF$_2$ with the simple rutile-type structure. We also report the results of our ab-initio calculations that reproduce qualitatively the observed NTE in ZnF$_2$.
\begin{figure}
\resizebox{0.5\textwidth}{!}{\includegraphics{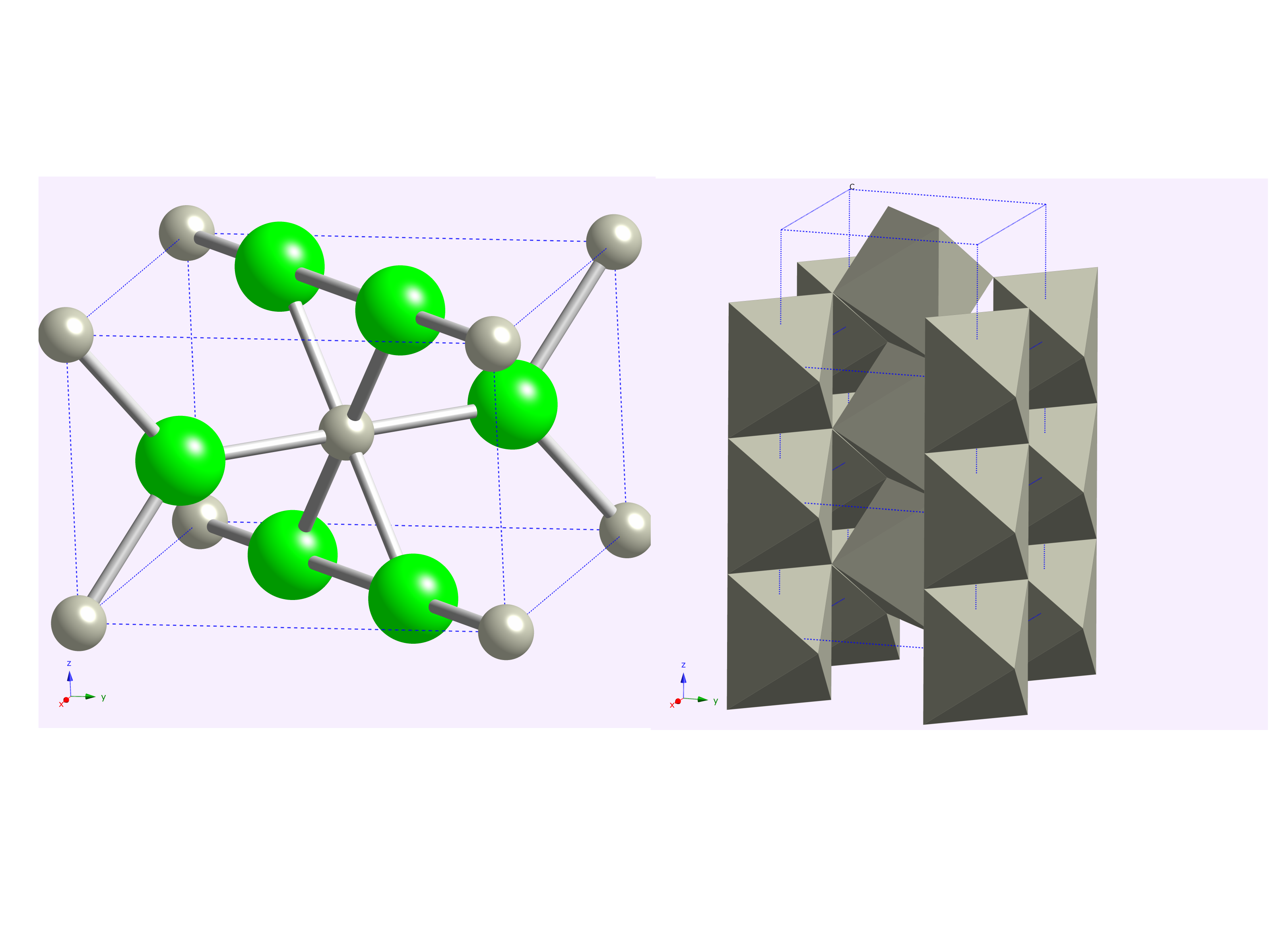}}

\caption {(Color online) (Left panel) Rutile-type crystal structure of ZnF$_2$. The grey spheres are Zn and the green spheres represent F atoms. (Right panel) The stacking of ZnF$_6$ octahedra in ZnF$_2$.  }
\label{structure}
\end{figure}
\begin{figure}
\resizebox{0.45\textwidth}{!}{\includegraphics{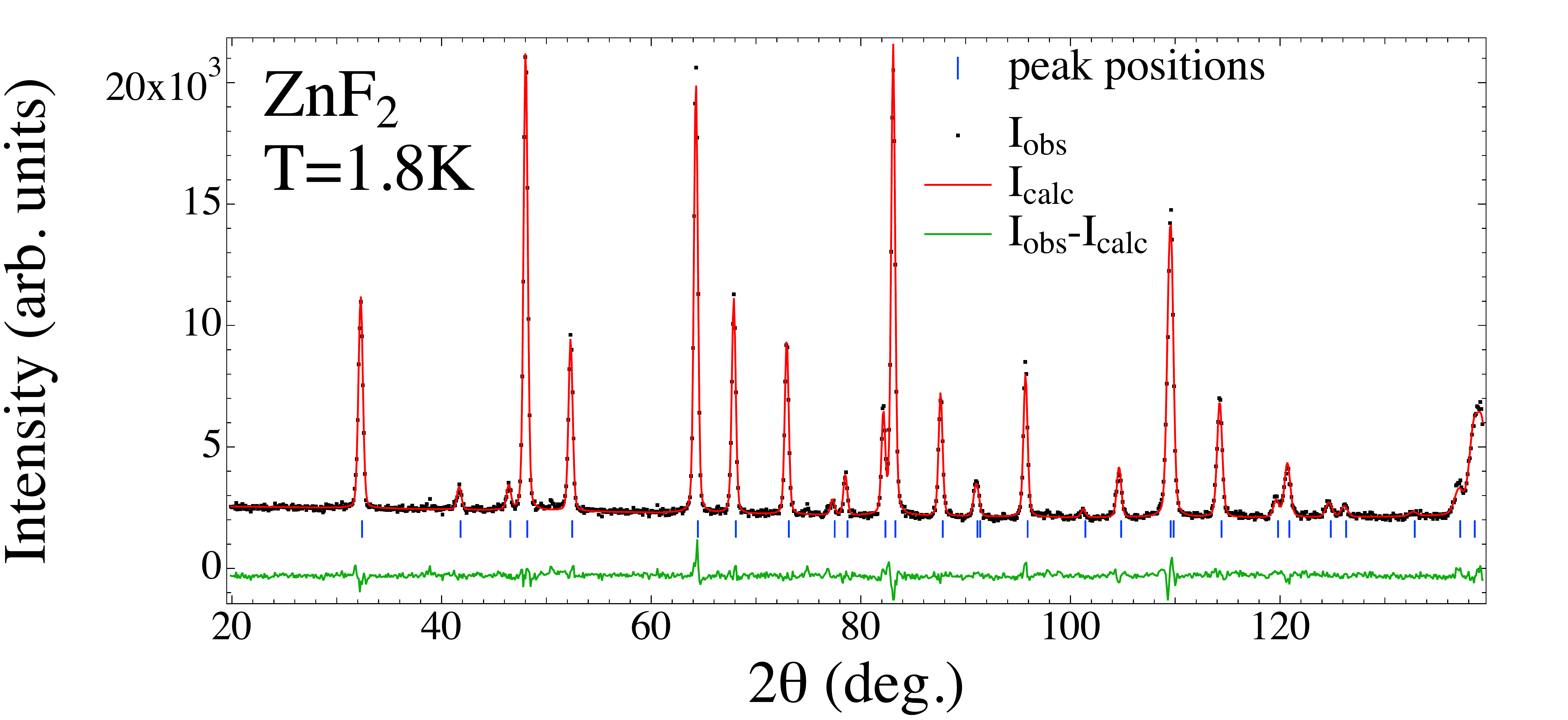}}
\caption {(Color online) Rietveld refinement of the  ZnF$_2$  crystal structure at T = 1.8 K. }
\label{znf2refinement}
\end{figure}
\begin{figure}
\resizebox{0.45\textwidth}{!}{\includegraphics{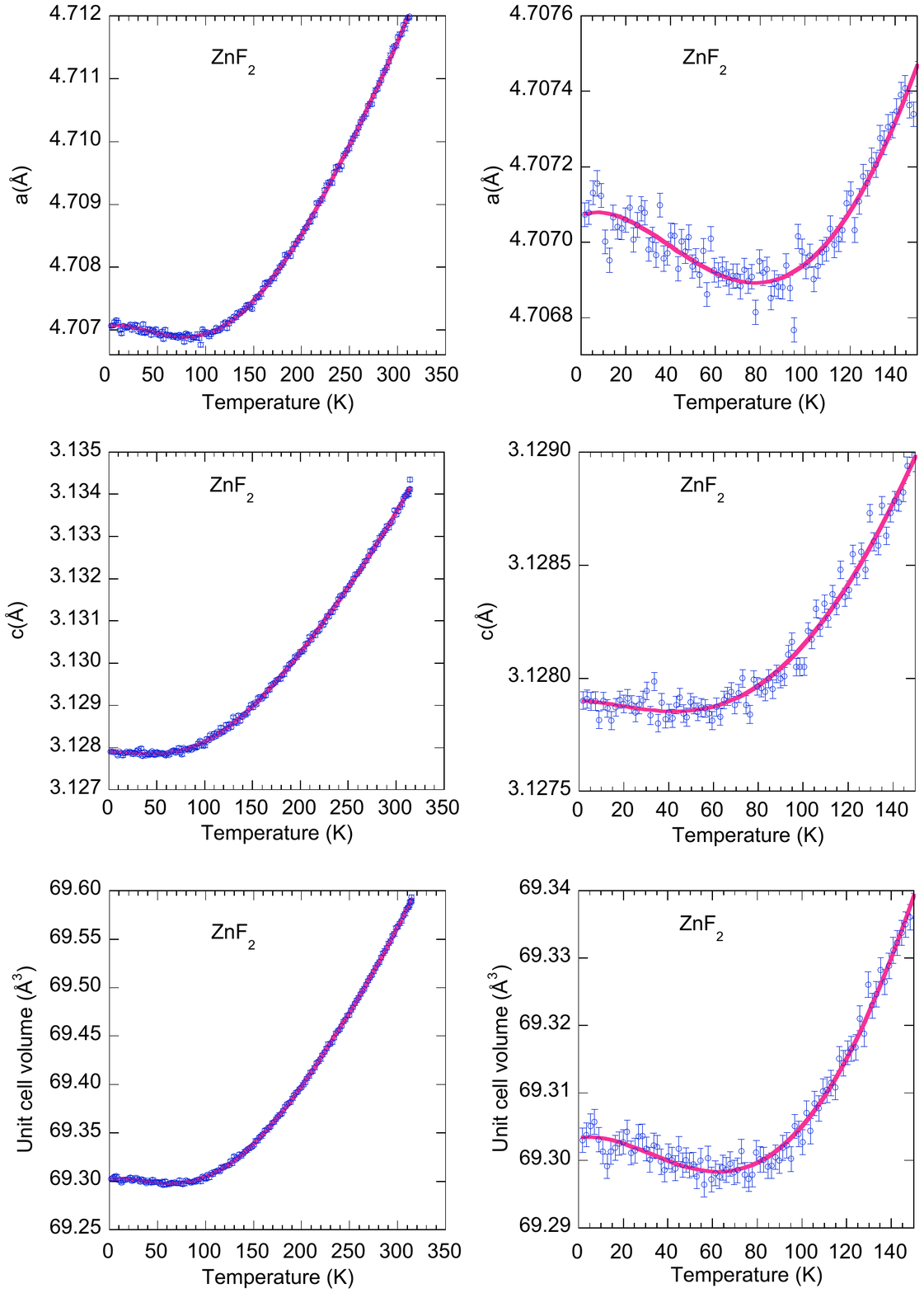}}
\caption {(Color online) Temperature variation of the lattice parameters  $a$, $c$, and the unit cell volume $V$  of ZnF$_2$ plotted on the left panels. The red curves in these figures represent the lattice parameter and the unit cell volume obtained by fitting the data by fifth order polynomials. On the right panels only the low temperature data are shown. }
\label{znlattice}
\end{figure}

\begin{figure}
\resizebox{0.5\textwidth}{!}{\includegraphics{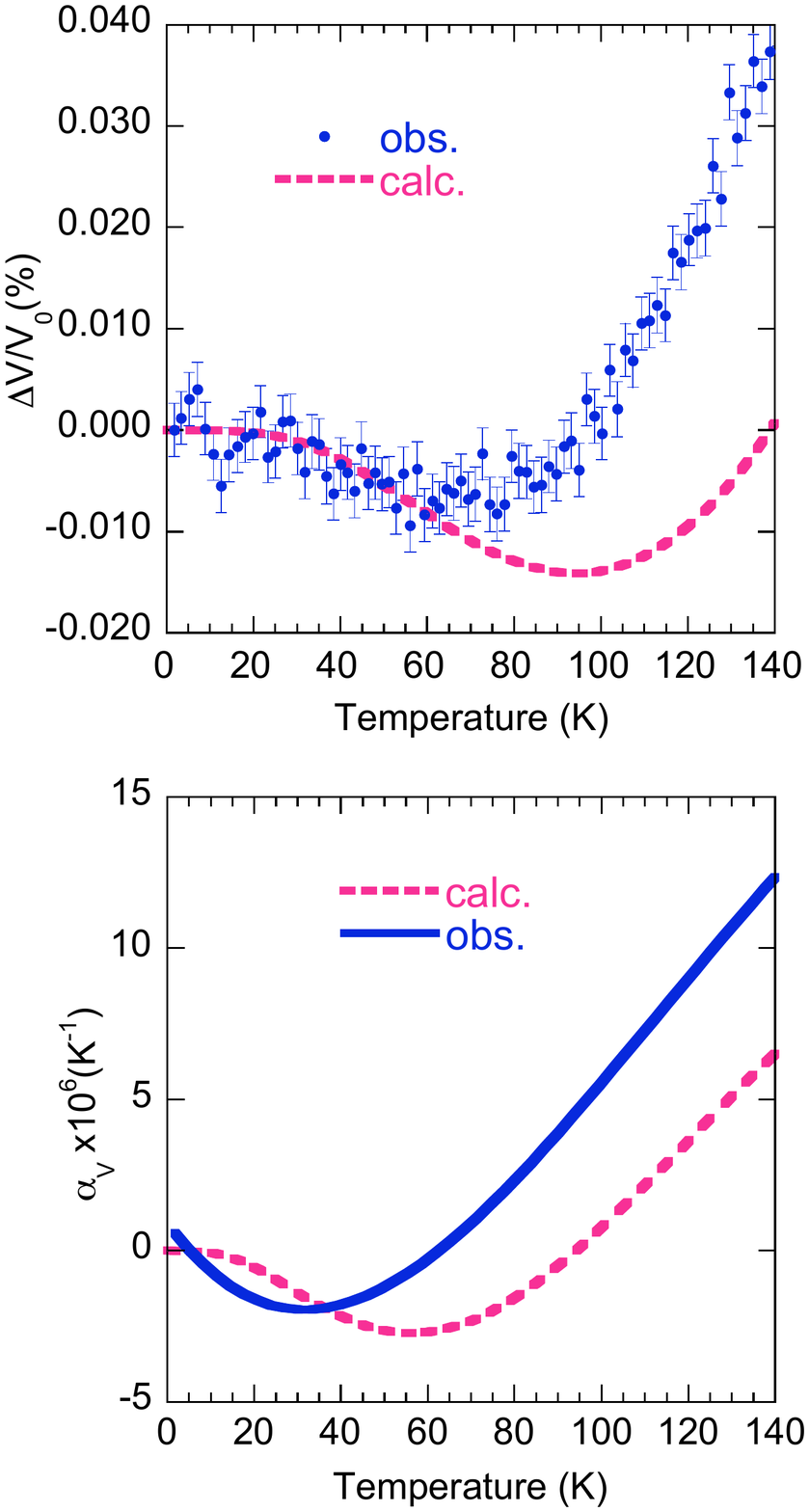}}
\caption {(Color online) (Upper panel)Temperature variation of $\Delta V/V_0$. The blue circles are the data points and the dotted red curve is the DFT calculation of $\Delta V/V_0$.(Lower panel) Temperature variation of the thermal expansion coefficient $\alpha_V = \frac{1}{V}\frac{dV}{dT}$. The blue and the dotted red curves correspond to the observed and calculated volume thermal expansion coefficient $\alpha_V$. The experimental values were determined from the differentiation of the fitted data with a fifth-order polynomial function.}
\label{comparison}
\end{figure}

\begin{figure}
\resizebox{0.5\textwidth}{!}{\includegraphics{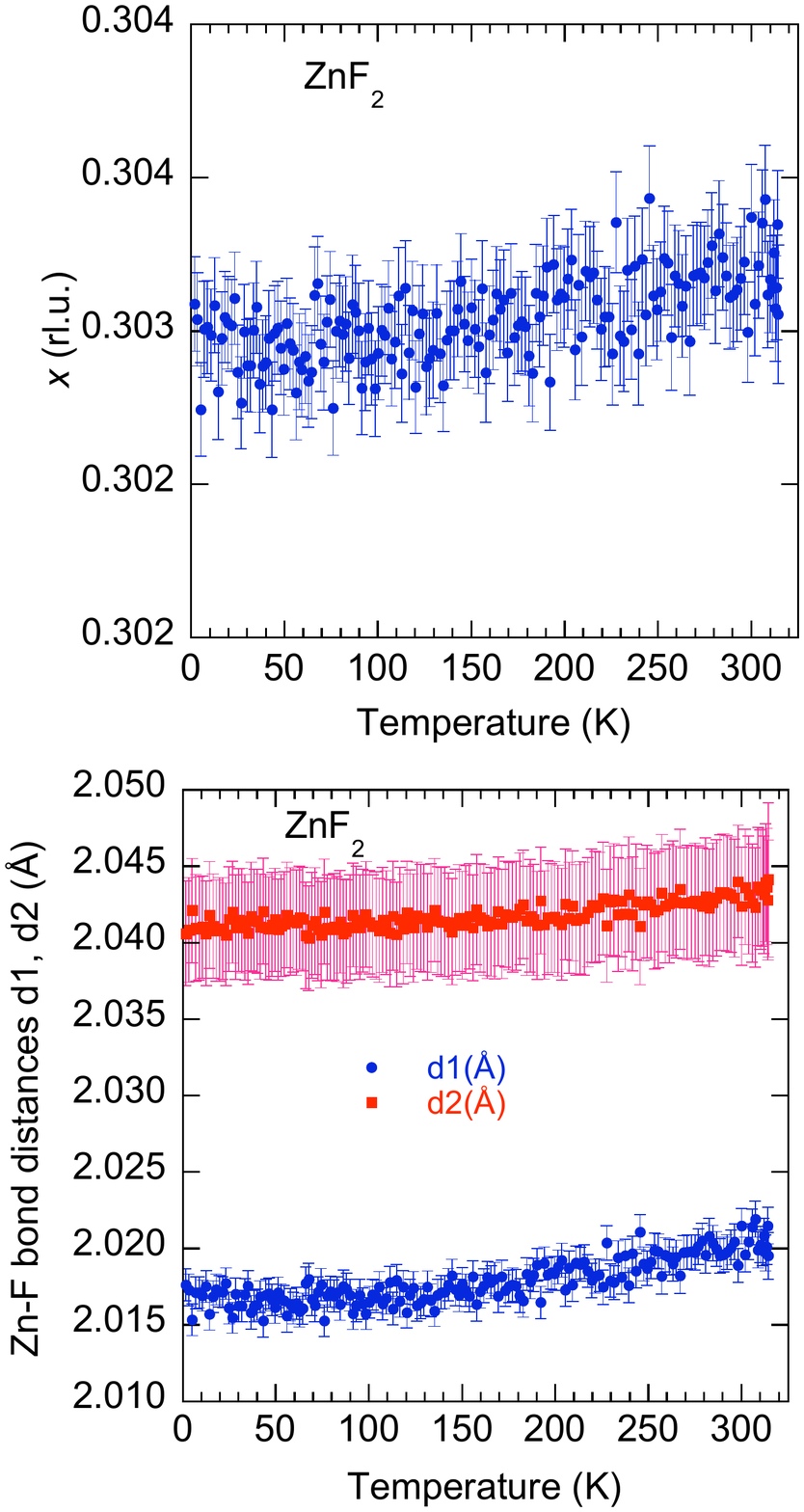}}
\caption {(Color online) Temperature variation of the positional parameter  $x$ of the F atom and the two Zn-F bond distances $d1$ and $d2$ of ZnF$_2$.  }
\label{bonds}
\end{figure}

The transition-metal difluorides MF$_2$ (M = V, Cr, Mn, Fe, Ni, Cu, Zn) with the rutile-type or distorted rutile-type crystal structure form an important class of materials with interesting magnetic and magneto-optic properties. In order to entangle magnetic effects from the lattice effects, the last of this series viz. the non-magnetic ZnF$_2$ has often been used to study the background lattice effects. However even this non-magnetic solid showed anomalies in elastic constants at low temperatures not expected for ZnF$_2$ with no temperature-induced phase transition. The observed softening of $C_{44}$ and $C_s = (C_{11} - C_{12})/2$ at low temperature in ZnF$_2$ had been interpreted as an incipient ferroelectric transition \cite{rimai77,boccara68} but has also been contested \cite{steiner94,vassiliou86}.

Figure \ref{structure} shows schematically the rutile-type crystal strucure of ZnF$_2$ that crystallizes with the space group $D_{4h}^{14}$ or $P4/mnm$. The unit cell is tetragonal with lattice parameter $a =  4.7034$, $c = 3.1335$ {\AA} at $T = 296$ K and it contains two formula units $Z= 2$. The two Zn$^{2+}$ ions are located at positions $(0,0,0)$ and $(1/2,1/2,1/2)$ whereas four F$^{-}$ ions are located at $(x,x,0)$, $(1-x,1-x,0)$, $(1/2-x,1/2+x,1/2)$, $(1/2+x,1/2-x,1/2)$ with the positional parameter $x = 0.303$. The Zn$^{2+}$ ions are surrounded by six F$^{-}$ ions to form slightly distorted octahedra. The octahedra are edge linked along the $c$-axis and corner-linked along $<110>$ crystallographic directions.

Neutron diffraction experiments were done on ZnF$_2$ on the high intensity powder diffractometer D20 of the Institute Laue-Langevin in Grenoble. The $115$ reflection from a Ge monochromator at a high take-off angle of $118^{\circ}$ gave neutron wavelength of 1.868 {\AA}.  Approximately 5 g ZnF$_2$  powder samples was placed inside an $8$ mm diameter vanadium can, which was fixed to the sample stick of a standard $^4$He cryostat. We have measured the diffraction intensities from ZnF$_2$ as a temperature in the range $1.8 - 320$ K. The Rietveld refinement\cite{rietveld69} of the diffraction data was done by the Fullprof program\cite{rodriguez10}. The refinement results from ZnF$_2$ at T = 1.8 K is shown in Fig. \ref{znf2refinement}.
The agreement factors R (not corrected for background) for pattern of this refinement were $R_p = 3.13$ and $R_{wp} = 4.13$. The corresponding conventional Rietveld R-factors were $R_p = 14.0$ and $R_{wp}= 10.5$. The goodness of the fit as given by $\chi^2$ was $\chi^2 = 2.07$.

Fig. \ref{znlattice} shows the temperature variation of the lattice parameters $a$, $c$, and the unit cell volume $V$  of ZnF$_2$ on the left panels. The red curves in these figures represent the lattice parameters and the unit cell volume obtained by fitting the data by  fifth degree polynomials. Attempts to fit the data by Debye or Einstein functions in Gr\"uneisen approximation failed for $a$ and $V$ because of the negative thermal expansion at low temperaure. Fifth degree polynomial function fit the low temperature successfuly. On the right panels only the low temperature data are shown. The lattice parameter $a$ and the unit cell volume $V$ exhibit minima at about 75 K.

The upper panel of Fig. \ref{comparison} shows the experimental normalized volume change given by 
\begin{equation}
\frac{\Delta V}{V_0}=\frac{V(T)-V_0}{V_0}
\label{VvsT}
\end{equation} 
where $V(T)$  is the unit cell volume at temperature $T$ and $V_0$ is the
volume at $T=0$ along with the values calculated within the density functional theory framework (DFT). The lower panel of Fig. \ref{comparison} shows the experimental and calculated volume thermal expansion coefficients $\alpha_V$ in the low temperature range given by
\begin{equation}
\alpha_V(T)=\frac{1}{V}\frac{dV}{dT}.
\label{alphav}
\end{equation}

In order to check whether there exists any indication of incipient ferroelectric phase transition \cite{rimai77,boccara68,steiner94,vassiliou86} in ZnF$_2$ at low temperature we refined the neutron powder diffraction data by the Rietveld method and determined the positional parameter $x$ and the two Zn-F bond distances $d_1$ and $d_2$ as a function of temperature. Fig. \ref{bonds} shows these quantities. The absence of any anomalies suggest that apart from the  negative thermal expansion (NTE) no further structural changes take place in ZnF$_2$ low temperature.

We have done calculations to check whether we can reproduce NTE in ZnF$_2$ using first-principles DFT. Since DFT is a $T= 0$ K approach, the finite temperature dependence is introduced within the framework of phonons by applying the so-called quasiharmonic approximation. The anharmonic effects are included uniquely via the volume dependence of the phonon frequencies. For a set of volumes around the equilibrium one, the procedure consists of evaluating the total Helmhotz free energy given by

\begin{equation}
F(V,T) = E_{el}(V) + F_{ph}(V,T)
\label{helmhotz}
\end{equation}
where E$_{el}$(V) and F$_{ph}(V,T)$ are the total ground-state temperature-free energy at constant volume as obtained directly from DFT and the phonon free energy extracted from subsequent
lattice dynamical calculations for each volume, respectively. The free energy $F(V,T)$ can be then used to evaluate thermodynamics of the material under study.
   
The starting geometry for the calculations was the experimentally refined
ZnF$_2$ structure~\cite{rimai77}. Relaxed geometries, total energies, phonon frequencies and volume-dependent phonon free energies were obtained using similar computational procedure described previously~\cite{compproc1,compproc2,compproc3}. 

The underlying mechanism of NTE in ZnF$_2$ seems to be very similar to that for tetrahedral semiconductors like Si, Ge, ZnS etc. with diamond and zincblende structures \cite{gibbons58,sparks67,biernacki89,barrera05}. It is the excitations at low temperatures of the low-energy phonon modes  with negative Gr\"uneisen parameters that are responsible for NTE in ZnF$_2$. These modes are likely connected with the rigid-mode vibrations of ZnF$_6$ octahedra and their linkage along the a axis shown in the right panel of Fig. \ref{structure}.

In conclusion we have done neutron diffraction study of the temperature dependence of the crystal structure of the simple non-magnetic or diamagnetic transition metal difluoride ZnF$_2$ with rutile structure and have discovered negative thermal expansion (NTE) at low temperature. Our first principle calculations reproduces qualitatively this experimental result.


\begin{thebibliography}{99}
\bibitem{mary96}T.A. Mary, J.S.O. Evans, T. Vogt and A.W. Sleight, Science {\bf 272}, 90 (1996).
\bibitem{evans96}J.S.O. Evans, T.A. Mary, T. Vogt, M.A. Subramanian and A.W. Sleight, Chem. Mater. {\bf 8}, 2809 (1996).
\bibitem{sleight98}A.W. Sleight, Ann. Rev. Mater. Sci. {\bf 28}, 29 (1998).
\bibitem{evans99}J.S.O. Evans, J. Chem. Soc. Dalton Trans. 3317 (1999).
\bibitem{gibbons58}D.F. Gibbons, Phys. Rev. {\bf 112}, 136 (1958).
\bibitem{sparks67}P.W. Sparks, and C.A. Swenson, Phys. Rev. {\bf 163}, 163 (1967).
\bibitem{biernacki89}S.Biernacki, and M. Scheffler, Phys. Rev. Lett. {\bf 63}, 290 (1989).
\bibitem{barrera05}G.D. Barrera, J.A.O. Bruno, T.H.K. Barron and N.L. Allan, J. Phys.: Condens. Matter {\bf 17}, R217 (2005).
\bibitem{rimai77}D.S. Rimai, Phys. Rev. B {\bf 16}, 4069 (1977).
\bibitem{boccara68}N. Boccara, Ann. Phys. (N.Y.) {\bf 47}, 40 (1968).
\bibitem{steiner94}M. Steiner, W. Potzel, M. K\"offerline, H. Karzel, W. Schiessl, G.M. Kalvius, D.W. Mitchell, N. Sahoo, H.H. Klauss, T.P. Das, R.S. Feigelson and G. Schmidt, Phys. Rev. B {\bf 50}, 13355 (1994).
\bibitem{vassiliou86}J.K. Vassiliou, J. Appl. Phys. {\bf 59}, 1125 (1986).
\bibitem{rietveld69}H.M. Rietveld, J. Appl. Cryst. {\bf 2}, 65 (1969).
\bibitem{rodriguez10}J. Rodriguez-Carvajal, FULLPROF, a Rietveld and pattern matching and analysis program version 2010, LLB, CEA-CNRS, France [http://www.ill.eu/sites/fullprof/]
\bibitem{compproc1} M. Zbiri, H. Mutka, M. R. Johnson, H. Schober and C. Payen, Phys. Rev. B {\bf 81}, 104414 (2010).
\bibitem{compproc2} The exchange-correlation contribution has been described presently by the local density approximation (LDA) based on the Ceperly-Alder parametrization by Perdew and Zunger~\cite{pz}.
\bibitem{compproc3} In order to determine accurately all the force constants, the supercell approach was used for lattice dynamics calculations. A tetragonal supercell ($2\times a$, $2\times b$, $3\times c$) was constructed from the relaxed geometry containing 24 formula-units (72 atoms). Total energies and Hellmann-Feynman forces were calculated for 8 structures resulting  from individual displacements of the symmetry inequivalent atoms in the supercell, along with the inequivalent cartesian directions ($\pm$x, $\pm$y and $\pm$z).
\bibitem{pz} J.P. Perdew and A. Zunger, Phys. Rev. B {\bf B23}, 5048 (1982).

\end{thebibliography}
\end{document}